\documentclass[epjST]{svjour}
\usepackage{graphics} 

\def\pr{\prime}

\def\be{\begin{equation}}
\def\lan{\left\langle}
\def\ran{\right\rangle}
\def\ee{\end{equation}}
\def\barr{\begin{array}}
\def\earr{\end{array}}

\def\l{\left}
\def\r{\right}
\def\f{\frac}
\def\dis{\displaystyle}
\def\ed{\end{document}}
\def\la{\lambda}

\def\cc{{\cal C}}

\def\baa{{\mbox{\boldmath $\alpha$}}}

\begin{document}
\title{Quadrupole properties of the eight $SU(3)$ algebras in $(sdgi)$ space}
\author{R. Sahu\inst{1}\fnmsep\thanks{\email{rankasahu@gmail.com}} 
\and V.K.B. Kota\inst{2} \and P.C. Srivastava\inst{3}}
\institute{National Institute of Science and Technology,
Berhampur 761008, India \and Physical Research Laboratory, Ahmedabad 380 009, 
India \and Department of Physics, Indian Institute of Technology, Roorkee 
247 667, India}
\abstract{With nucleons occupying an oscillator shell $\eta$, there are
$2^{[\f{\eta}{2}]}$ number of $SU(3)$ algebras; $[\f{\eta}{2}]$ is the integer
part of $\eta/2$. Analyzing the first non trivial situation with four $SU(3)$
algebras in $(sdg)$ space, demonstrated recently is that they generate quite
different quadrupole properties though they all generate the same spectrum. More
complex situation is with eight $SU(3)$ algebras in $(sdgi)$ space. In the
present work, quadrupole properties generated by these eight algebras are
analyzed first using the more analytically tractable interacting
boson model. In addition, shell model and the closely related deformed 
shell model are used with three examples of nucleons in $sdgi$ space. 
It is found that in general six of the $SU(3)$
algebras generate prolate shape and two oblate shape. Out of all these, one of the $SU(3)$ algebra generates quite small quadrupole moments for the
low-lying states.} 
%
\maketitle
\section{Introduction}
\label{intro}

Shell model (SM) and the interacting boson model (IBM) admit $SU(3)$  algebra
generating rotational spectra in nuclei. Going beyond the introduction of
$SU(3)$ in SM by Elliott \cite{Ell-1,Ell-2} that is applicable only to light
nuclei, the scope of $SU(3)$ in nuclei is enlarged, spreading its applicability
all across the periodic chart,  by the developments in the pseudo-$SU(3)$ model
\cite{jpd-1,jpd-2}, $SU(3)$ of fermion dynamical symmetry model \cite{fdsm},
proxy-$SU(3)$ model \cite{Bona} and the $Sp(6,R)$ model containing $SU(3)$
\cite{Rowe-1,Rowe-2,Kristina} all within the shell model on one hand and by
various extended IBM's such as IBM-2,3,4 \cite{Iac-87}, $sdg$IBM
\cite{YK-1,YK-2}, $sdpf$IBM \cite{Long,Iac-1}, IBFM (interacting boson-fermion
model) and IBFFM (interacting boson-fermion-fermion model) \cite{BK,Iac-2,KU} on
the other. In addition, there are the algebraic cluster model
\cite{Cseh-1,Cseh-2} and many other models that employ $SU(3)$ symmetry.  A new
paradigm in the applications of $SU(3)$ in nuclei is the recent recognition that
for nucleons in   a given oscillator shell $\eta$, there will be multiple
$SU(3)$ algebras \cite{Ko-17,Ko-19}. For a recent overview of $SU(3)$ symmetry in atomic nuclei, see \cite{kota-su3}.

Simplest example of two $SU(3)$ algebras in the $\eta=2$ shell is well known
\cite{Parikh,Iac-87} and this is the situation with $(2s1d)$ shell in SM and
also $sd$IBM. The prolate shape generated by one of the $SU(3)$ algebras and
oblate shape by the other are found to play an important role in quantum phase
transition studies (QPT) \cite{RMP}. However,  the first non-trivial example of
four $SU(3)$ algebras in $sdg$ space was analyzed only recently \cite{Ko-19}.
Used in this study are $(sdg)^{6p}$, $(sdg)^{6p,2n}$ and $(sdg)^{6p,
6n}$ examples in SM and a 10 boson system in $sdg$IBM with $sdg$IBM giving
simple analytical results in the large boson number limit (note that $p$ stands
for protons and $n$ for neutrons).  Let us mention that in the $(sdg)^{6p ,
6n}$ example, as the shell model matrix dimensions are very large, employed is the
deformed shell model (DSM) based on Hartree-Fock single particle states
\cite{RS}.  It is found that the four $SU(3)$ algebras in the $(sdg)$ space
exhibit quite different properties with regard to the quadrupole collectivity as
brought out by the quadrupole moments $Q(J)$ and $B(E2)$'s in the ground $K=0$
band in the even-even nuclei studied. The general structure observed in the
$sdg$ examples is that one of the $SU(3)$ algebras generates prolate shape, one
oblate and the other two also generate  prolate shape but one of them gives very
small quadrupole moments for the low-lying levels. Thus, with multiple $SU(3)$ algebras, it is possible to have rotational spectra (for example, a $K=0$ band with $J=0,2,4,\ldots$) with very small quadrupole transition matrix elements. Our purpose in the present
paper is to analyze the more complex situation of $(sdgi)$ space that admits
eight $SU(3)$ algebras. Now we will give a preview.

In Section 2, first introduced are the multiple $SU(3)$ algebras for bosons in an
oscillator shell $\eta$ and then restricting to $(sdgi)$ space, the structure of
the ground $K=0$ bands generated by the eight algebras in the $(sdgi)$ space is
studied. Section 3 gives the formulas for quadrupole moments ($Q(L)$) and
$B(E2)$'s for the ground $K=0$ band and using them results generated for the
quadrupole properties by the eight $SU(3)$ algebras are presented. In Section 4,
results are presented for the  shell model $(sdgi)^{6p}$, $(sdgi)^{6p,6n}$ and $(sdgi)^{12p,6n}$ systems using SM codes for the first and DSM for all the three examples. Finally, Section 5 gives conclusions.

\section{Eight $SU(3)$ algebras in $(sdgi)$IBM and structure of K=0 bands}
\label{sec-1}

Given a oscillator shell $\eta$, occupied by say $N$ number of bosons, there
will be multiple $SU(3)$ algebras generated by the angular momentum operators
$L^1_q$ and quadrupole moment operators $Q^2_\mu(\baa)$ given by \cite{Ko-19}
\be
\barr{l}
L^1_q = \dis\sum_\ell \dis\sqrt{\dis\frac{\ell(\ell+1)(2\ell+1)}{3}}
\l(b^\dagger_\ell \tilde{b}_\ell\r)^1_q \;,\\
Q^2_\mu(\baa)= -(2\eta+3)\,\dis\sum_\ell
\dis\sqrt{\dis\frac{\ell(\ell+1)(2\ell+1)}{5(2\ell+3)(2\ell-1)}}
\l(b^\dagger_\ell \tilde{b}_\ell\r)^2_\mu \\
+ \dis\sum_{\ell <
\eta}\,\alpha_{\ell,\ell+2}\;\dis\sqrt{\dis\frac{6(\ell+1)(\ell+2)(\eta-\ell)
(\eta+\ell+3)}{5(2\ell+3)}}\l[\l(b^\dagger_\ell \tilde{b}_{\ell+2}\r)^2_\mu +
\l(b^\dagger_{\ell+2} \tilde{b}_{\ell}\r)^2_\mu \r]\;; \\
\baa=(\alpha_{0,2}, \alpha_{2,4}, \ldots, \alpha_{\eta-2,\eta})\;\;
\mbox{for}\;\;\eta\;\;\mbox{even}\;,\\
\baa=(\alpha_{1,3}, \alpha_{3,5}, \ldots, \alpha_{\eta-2,\eta})\;\;
\mbox{for}\;\;\eta\;\;\mbox{odd}\;,\\
\baa=(\pm 1, \pm 1, \ldots)\;.
\earr \label{eq.epj-1}
\ee
Thus, for each choice of $\baa$ there is a $SU^{\baa}(3)$ algebra. Clearly,
given a $\eta$, the number of $SU(3)$ algebras is $2^{\l[\f{\eta}{2}\r]}$ where
$\l[\f{\eta}{2}\r]$ is the integer part of $\eta/2$. Then, there will be two
$SU(3)$ algebras for $\eta=2$ shell ($sd$ space), four in $\eta=4$ shell ($sdg$
space), eight in $\eta=6$ shell ($sdgi$ space) and so on (if we have two
different shells as in $sdpf$IBM or bosons with internal degrees of freedom as
in IBM-2,3,4 models or Bose-Fermi systems, the number of $SU(3)$ algebras will
be in general many more and these are not considered in the present paper).
Standard choice for the $\baa$ as considered by Elliott and used in IBM studies
is $\alpha_{\ell , \ell +2}=+1$ independent of $\ell$ value. In this article we
will focus on  $(sdgi)$ space and consider the quadrupole properties generated 
by the eight $SU(3)$ algebras in this space for a $N$ boson system. With $SU(3)$
operating, the $(sdgi)^N$ space decomposes into irreducible subspaces labeled by
the irreducible representations (irreps) of $SU(3)$ and they are denoted by
$(\la \mu)$. The decomposition of $(\la \mu)$'s  into $L$ is well known giving the
$K$ quantum number. For example, a $(\la , 0)$ irrep gives $K=0$ with L=$\la$,
$\la -2$, $\ldots$, $0$ or $1$. Similarly, a $(\la , 2)$ irrep gives $K=0$ and
$K=2$ bands, a $(\la , 4)$ irrep gives $K=0$, $2$ and $4$ bands and so on.

In $sdgi$IBM with $\eta=6$, $(\baa)= (\alpha_{02}, \alpha_{24},
\alpha_{46})$ and with this there will be eight $SU^{ (\alpha_{02}, \alpha_{24},
\alpha_{46})}(3)$ algebras. With $b^\dagger_{0}=s^\dagger$, 
$b^\dagger_{2m}=d^\dagger_m$, $b^\dagger_{4m}=g^\dagger_m$, 
$b^\dagger_{6m}=i^\dagger_m$ and $\tilde{b}_{\ell m} = (-1)^{\ell -m} b_{\ell,
-m}$, the eight algebras are generated by the operators,
\be
\barr{l}
L^1_q = \dis\sqrt{10}  \l(b^\dagger_2 \tilde{b}_2\r)^1_q + 2\dis\sqrt{15}  
\l(b^\dagger_4 \tilde{b}_4\r)^1_q + \dis\sqrt{182} \l(b^\dagger_6 \tilde{b}_6
\r)^1_q \;,\\
Q^2_\mu(\alpha_{02},\alpha_{24},\alpha_{46}) =
\dis\sum_{\ell , \ell^\prime =0,2,4}\;t_{\ell , \ell^\prime}(\alpha_{02}, 
\alpha_{24}, \alpha_{46})\;
\l(b^\dagger_\ell \tilde{b}_{\ell^\prime}\r)^2_\mu\;;\\
t_{2,2} = -15\;\dis\sqrt{\dis\frac{2}{7}}\;,\;\;t_{4,4} =
-\dis\frac{90}{\dis\sqrt{77}}\;,\;\; t_{6,6} = -3\dis\sqrt{\dis\f{182}{11}}\;,
\;\;t_{0,2} = t_{2,0} = \alpha_{02}\,6 \dis\sqrt{\dis\frac{6}{5}} \;,\\
t_{2,4} = t_{4,2} =  \alpha_{24}\,12\dis\sqrt{ \dis\f{22}{35}}\;,\;\;
t_{4,6} = t_{6,4} = \alpha_{46}\,6\dis\sqrt{ \dis\f{26}{11}} \;.
\earr \label{eq.epj-2}
\ee
Note that $(\alpha_{02}, \alpha_{24}, \alpha_{46})$ = $(\alpha_{sd},\alpha_{dg},
\alpha_{gi})$ = $(+,+,+)$, $(+,+,-)$, $(+,-,+)$, $(+,-,-)$, $(-,+,+)$,
$(-,+,-)$, $(-,-,+)$ and $(-,-,-)$ where $+$ stands for $+1$ and $-$ stands for
$-1$.  Correspondingly, there are eight quadrupole-quadrupole Hamiltonians 
\be
H_Q^{(\alpha_{02},\alpha_{24}, \alpha_{46})} = -(1/4)\, Q^2(\alpha_{02},
\alpha_{24},\alpha_{46})  \cdot Q^2(\alpha_{02},\alpha_{24},\alpha_{46}) \;.
\label{eq.epj-3}
\ee
As $H_Q^{(\alpha_{02},\alpha_{24},\alpha_{46})} =-\cc_2(SU^{(\alpha_{02} ,
\alpha_{24}, \alpha_{46})}(3)) +(3/4)L^2$, their eigenvalues over an $SU(3)$
state $\l|(\la \mu)KL\ran$ are $-[\la^2 +\mu^2 + \la \mu +3(\la  + \mu)] +
\f{3}{4}L(L+1)$. Therefore,  all these eight $H_Q$'s generate the same spectrum
and the ground band belongs to the $SU(3)$ irrep $(6N,0)$ for a $N$ boson
system. This is generated by the intrinsic state
\be
\barr{l}
\l|N; x_0,x_2,x_4,x_6\ran = (N!)^{-1/2} \l(x_0 s^\dagger_0 + x_2
d^\dagger_0 +x_4 g^\dagger_0 +x_6 i^\dagger_0\r)^N \l|0\ran \\
\Rightarrow \l|N; \beta_2, \beta_4, \beta_6\ran = 
\l[N!\; (1+\beta_2^2 + \beta_4^2+\beta_6^2)^N \r]^{-1/2} 
\l(s^\dagger_0 + \beta_2
d^\dagger_0 +\beta_4 g^\dagger_0 +\beta_6 i^\dagger_0\r)^N \l|0\ran\;.
\earr \label{eq.epj-4}
\ee
Note that $x_0^2+x_2^2+x_4^2+x_6^2=1$ and the choice $x_0=1/x$, $x_2=\beta_2/x$,
$x_4=\beta_4/x$ and $x_6=\beta_6/x$ gives the second form above; without loss of
generality, choosing $x$ to be positive, we have  $x=(1+\beta_2^2 + \beta_4^2
+ \beta_6^2)^{1/2}$. Now, the energy functional is given by
\be
\barr{l} 
E_{SU^{(\alpha_{02},\alpha_{24}, \alpha_{46})}(3)}\l(N; \beta_2 , \beta_4 , 
\beta_6 \r) = 
\lan N; \beta_2, \beta_4, \beta_6
\mid H_Q^{(\alpha_{02},\alpha_{24}, \alpha_{46})} \mid N; \beta_2, \beta_4, 
\beta_6 \ran \\
= -\frac{9}{29645 \left(1+ \beta_2^2+ \beta_4^2+ \beta_6^2\right)^2} \;
\l\{ 15125  \beta_2^4+2420 \beta_2^3 \left(7 \sqrt{30}  {\alpha_{02}}+4 
\sqrt{55}  \beta_4  {\alpha_{24}}\right) \r. \\
+22  \beta_2^2 \left(6468+5122
\beta_4^2+1225  \beta_6^2+1232 \sqrt{66} {\beta_4}  {\alpha_{02}}   
{\alpha_{24}}+1750 \sqrt{2}  {\beta_4}  {\beta_6}  {\alpha_{46}}\right) \\
+ 5 \left(2500  \beta_4^4+14700  \beta_4^2  \beta_6^2+2401  \beta_6^4
+ 7000 \sqrt{2}  \beta_4^3  {\beta_6}  {\alpha_{46}}+6860 \sqrt{2}  {\beta_4}
 \beta_6^3  {\alpha_{46}}\right) \\
+ 44\, {\beta_2} \left[ 7 \sqrt{15}  {\alpha_{02}} \left(50 \sqrt{2}  
\beta_4^2 + 49 \sqrt{2}  \beta_6^2 + 140  {\beta_4}  {\beta_6}  
{\alpha_{46}}\right) \r. \\
+ \l.\l. 4 \sqrt{55}  {\beta_4}  {\alpha_{24}} \left(50  \beta_4^2 + 49  
\beta_6^2+70 \sqrt{2}  {\beta_4}  {\beta_6}  {\alpha_{46}}\right)\right]\r\}
\earr \label{eq.epj-5}
\ee
Minimizing the energy functional with respect to $\beta_2$, $\beta_4$ and 
$\beta_6$ will give the parameters $(x_0,x_2,x_4,x_6)$ that define the intrinsic
state for the ground $K=0$ band for the eight $SU(3)$ algebras. For a given
$(\alpha_{02}, \alpha_{24}, \alpha_{46})$, the parameters $x_\ell$ defining 
the intrinsic state in Eq. (\ref{eq.epj-4}), after simplifying the results in \cite{kota-su3}, are given by
\be
x_0 = \dis\sqrt{\dis\f{33}{231}}\,,\;\; 
x_2 = \alpha_{02}\, \dis\sqrt{\dis\frac{110}{231}}\,,\;\; 
x_4 = \alpha_{02} \alpha_{24}\, \dis\sqrt{\dis\frac{72}{231}}\,,\;\;
x_6 = \alpha_{02} \alpha_{24} \alpha_{46}\,\dis\sqrt{\dis\frac{16}{231}} \;.
\label{eq.epj-6}
\ee
As seen from Eq. (\ref{eq.epj-6}), the intrinsic states for the ground $K=0$
band for the eight $SU(3)$'s differ only in the phases of $x_\ell$'s and this is
similar to the situation with $sdg$IBM \cite{Ko-19}. In addition, the intrinsic
state for $SU^{(+,+,+)}(3)$ has the same structure as the 
Nilsson orbit $\l[660\r]$ in
orbital space. Substituting these values of $x_\ell$'s in Eq. (\ref{eq.epj-5})
gives $E^0_{SU_{sdg}(3)}=-36N^2$  for all the eight $SU(3)$ algebras in
agreement with the large-$N$ limit result for the ground $SU(3)$ irrep $(6N,0)$.
However, the quadrupole properties (quadrupole moments and $B(E2)$'s) generated
by them, as obtained using a quadrupole transition operator, will be different.
We will turn to this now.

\section{Quadrupole properties from the eight $SU(3)$ algebras in $sdgi$IBM} 
\label{sec-2}

In order to study the quadrupole moments and $B(E2)$ values in the ground $K=0$ 
bands generated by the eight $SU^{(\alpha_{02} , \alpha_{24},  \alpha_{46})}(3)$
algebras in $sdgi$IBM, the $E2$ transition operator is chosen to be  
$$
T^{E2} = q_2\,Q^2_q(+,+,+)
$$ 
where $q_2$ is a parameter. As stated before, in IBM the
standard choice for the quadrupole operator is $Q^2_q(\baa)$ with $\alpha_{\ell
, \ell +2} = +1$ for all $\ell$. Now, as $Q^2_q(+,+,+)$ is a generator of
$SU^{(+,+,+)}(3)$, formulas for the quadrupole moments $Q(L)$ and $B(E2)$s along
the $K=0$ ground band follow easily from Eq. (10) given ahead as the
eigenstates obtained from $H_Q^{(+,+,+)}$ belong to $SU^{(+,+,+)}(3)$.
However, for the other $H_Q$ operators, the ground bands belong to the $(6N,0)$
irrep of the corresponding $SU^{(\alpha_{02} , \alpha_{24},  \alpha_{46})}(3)$
algebras. Therefore, the $T^{E2}$ chosen is no longer a generator of these
algebras. Hence, the simple $SU(3)$ formulas given by Eq. (10) are not
valid for these and we have to use a much more detailed $SU(3)$ algebra. Instead
of this, large-$N$ limit formulas (valid to order $1/N^2$) are used in the
analysis.

Starting with the intrinsic state given by Eq. (\ref{eq.epj-4}), it is easy to
construct the angular momentum projected states $\l|N;K=0,L,M\ran$. Using these,
formulas for the  quadrupole moments $Q(L)$ and $B(E2; L \rightarrow L-2)$ for
the ground band are derived by Kuyucak and Morrison \cite{KuMo-1} that are valid
for any one-body $T^{E2}$ operator. These
formulas, valid to order $1/N^2$, are
\be 
\barr{l}
Q(L)=\lan LL \mid Q^2_0 \mid LL\ran = \dis\frac{\lan LL\;20 \mid LL\ran}{
\dis\sqrt{2L+1}}\;\lan L \mid\mid Q^2 \mid\mid L\ran\;,\\
B(E2;L \rightarrow L-2)= \dis\frac{5}{16\pi}\;\f{\l|\lan L-2 \mid\mid Q^2 
\mid\mid L\ran\r|^2}{(2L+1)}\;;\\
\lan N;K=0,L_f \mid\mid Q^2 \mid\mid N;K=0,L_i\ran =
\l[N\dis\sqrt{(2L_i+1)}\r] \lan L_i 0\;\;20 \mid L_f,0\ran \;\times \\
\l[B_{00} +\frac{1}{N}\l(B_{00} - \dis\frac{B_{10}-3B_{00}}{a}\r) 
-\dis\frac{L_i(L_f+1)}{aN^2}\l\{B_{00}
+\dis\frac{F_1}{4a}\;\delta_{L_f, L_i} \r.\r. \\
\l.\l. - \dis\frac{F_2}{12a}\;\delta_{L_f, L_i+2}
\r\}\r]\;; \;\;\;\;L_f=L_i\;\;\mbox{or}\;\;L_f=L_i+2\\
B_{mn}=\dis\sum_{\ell^\pr , \ell}\;\l[\ell^\pr(\ell^\pr+1)\r]^m 
\l[\ell (\ell+1)\r]^n \lan \ell^\pr
0\;\ell 0 \mid 20\ran \; t_{\ell^\pr,\ell}\;x_{\ell^\pr} x_{\ell}\;, \\
F_1=B_{20}-B_{11}-10B_{10}+12B_{00},\;\;F_2=B_{20}-B_{11}
+6B_{10}-12B_{00}, \\
a=\dis\sum_{\ell} \ell (\ell+1) \l(x_{\ell}\r)^2 \;.
\earr \label{eq.epj-7}
\ee
Note that $\ell=0$, $2$, $4$ and $6$  for $sdgi$IBM and the $t_{\ell^\pr , \ell}$
are the coefficients in the $E2$ transition operator as given in Eq.
(\ref{eq.epj-2}). Using $T^{E2}$ given above,  the solutions for
$x_\ell$ given in Eq. (\ref{eq.epj-6}) and the formulas in Eq. (\ref{eq.epj-7}), calculated are $Q(L)$ and $B(E2; L \rightarrow
L-2)$ with $L=2$, $4$, $6$, $8$ and $10$ for a 15 boson system and the
results are given in Table 1. Let us mention that for $SU^{(+,+,+)}(3)$, exact $SU(3)$ formulas are given by Eq. (10) ahead with the replacements $\lambda = 6N$, $J=L$, $X_{eff}b^2=q_2$. Firstly, it can be verified that the results for
$SU^{(+,+,+)}(3)$ in Table 1 are essentially same as those from the exact $SU(3)$
formulas and this in turn is a good test of the formulas in Eq. (\ref{eq.epj-7}). More
importantly, it is seen that only $SU^{(-,-,+)}(3)$ and $SU^{(-,-,-)}(3)$
generate oblate shapes and all other six $SU(3)$'s generate prolate shapes.
Also, out of these six, only  $SU^{(+,+,+)}(3)$ and $SU^{(+,+,-)}(3)$  generate
large quadrupole moments and strong $B(E2)$'s. Similarly, $SU^{(+,-,-)}(3)$ generates quite small quadrupole moments (less by a factor 5 compared to the largest). Thus, the eight 
$SU(3)$ algebras generate quite different quadrupole properties for ground $K=0$
bands. For further understanding of the quadrupole properties generated by the 
eight $SU(3)$ algebras in $(sdgi)$ space, we will now consider shell model 
examples.

\begin{table}
\caption{Quadrupole moments $Q(L)$ and $B(E2; L \rightarrow L-2)$ values for the low-lying states in the
ground band for a 15 boson system generated by the eight $SU^{(\baa)}(3) $ algebras in  $sdgi$IBM. Note that in the table $Q(L)$ (in units of $q_2$) and $B(E2; L \rightarrow L-2)$ (in units of $(q_2)^2$) are given for $L=2$, $4$, $6$, $8$ and $10$ in columns 3 to 7 respectively.}
\label{table1}
\begin{tabular}{ccccccc}
\hline\noalign{\smallskip}
$\baa$ & $\;\;\;\;\;$ & \multicolumn{5}{c}{$J$} \\
& & 2 & 4 & 6 & 8 & 10 \\
\noalign{\smallskip}\hline\noalign{\smallskip}
$(+,+,+)$ & $Q$ & $-52$ & $-67$ & $-73$ & $-77$ & $-80$ \\
& $B(E2)$ & $666$ & $951$ & $1045$ & $1090$ & $1115$ \\
$(+,+,-)$ & $Q$ & $-43$ & $-55$ & $-61$ & $-64$ & $-66$ \\
& $B(E2)$ & $460$ & $654$ & $714$ & $739$ & $747$ \\
$(+,-,+)$ & $Q$ & $-18$ & $-24$ & $-26$ & $-28$ & $-29$ \\
& $B(E2)$ & $82$ & $116$ & $127$ & $131$ & $132$ \\
$(+,-,-)$ & $Q$ & $-10$ & $-12$ & $-14$ & $-15$ & $-16$ \\
& $B(E2)$ & $22$ & $30$ & $32$ & $31$ & $29$ \\
$(-,+,+)$ & $Q$ & $-21$ & $-27$ & $-30$ & $-32$ & $-34$ \\
& $B(E2)$ & $107$ & $154$ & $171$ & $181$ & $190$ \\
$(-,+,-)$ & $Q$ & $-12$ & $-16$ & $-18$ & $-19$ & $-20$ \\
& $B(E2)$ & $36$ & $51$ & $56$ & $58$ & $59$ \\
$(-,-,+)$ & $Q$ & $13$ & $16$ & $17$ & $17$ & $17$ \\
& $B(E2)$ & $41$ & $58$ & $64$ & $66$ & $66$ \\
$(-,-,-)$ & $Q$ & $22$ & $27$ & $30$ & $30$ & $30$ \\
& $B(E2)$ & $117$ & $167$ & $185$ & $195$ & $201$ \\
\noalign{\smallskip}\hline
\end{tabular}
\end{table}

\section{Shell model and deformed shell model analysis in $(sdgi)$ space}

\subsection{Preliminaries}

In the shell model with valence nucleons in $(sdgi)$ orbits, we have eight
$SU^{\baa}(3)$ algebras and the generators of these, in $LST$ coupling, follow
from Eq. (1) by replacing $(b^\dagger_{\ell_f} \tilde{b}_{\ell_i})^2_q$ by 2
$(a^\dagger_{\ell_f \f{1}{2} \f{1}{2}} \tilde{a}_{\ell_i \f{1}{2} \f{1}{2}})^{L_0=2,S_0=0,T_0=0}_q$. Then, we have 
\be
Q^2_q(\baa) = 2\dis\sum_{\ell_f , \ell_i} t_{\ell_f , \ell_i}(\baa)\;
\l(a^\dagger_{\ell_f \f{1}{2} \f{1}{2}}
\tilde{a}_{\ell_i \f{1}{2} \f{1}{2}}\r)^{2,0,0}_q\;.
\label{eq.epj-8}
\ee
Note that $t_{\ell_f , \ell_i}(\baa)$ are given in Eq. (2) and $\baa$ takes 8 values as given before. In the shell
model analysis used are the examples $(sdgi)^{6p}$, $(sdgi)^{6p,6n}$ and $(sdgi)^{12p,6n}$ systems giving the lowest $SU(3)$ irreps to be $(30,0)$,
$(60,0)$ and $(78,0)$ respectively \cite{Ko-hw}. Again, choosing 
$$
H_Q^{(\baa)} = -(1/4) Q^2(\baa) \cdot Q^2(\baa)\;,
$$
for the eight $SU(3)$ algebras studied are the energies of the
yrast ($K=0$ band) levels, quadrupole moments $Q_2(J)$ of these levels and the
$B(E2)$'s along the yrast line for $J$ up to 10. Used for this purpose are
DSM and also the Antoine shell model code \cite{Anton}. Note that DSM brings out
shape information in a transparent manner and also it is useful for larger
particle numbers where SM calculations are impractical \cite{RS}. For easy reference, in Appendix A given are the formulas for the single particle energies (spe) and two-body matrix elements (TBME) defining the $Q.Q$ operators and these are the inputs for both SM and DSM calculations.  

In the SM (also DSM)
studies, the $E2$ transition operator is taken to be 
\be
T^{E2} = \l[e^p_{eff}\,Q^2_q(-,-,-;p) + e^n_{eff}\,Q^2_q(-,-,-;n)\r]\,  b^2
\label{eq.e2op}
\ee
where $b$ is
the oscillator length parameter and $e^p_{eff}$ and $e^n_{eff}$ are proton and
neutron effective charges. This choice follows from the fact that in SM it is standard to use $\baa=(\alpha_{sd}, \alpha_{dg}, \alpha_{gi}) = (-,-,-)$; see for example \cite{Bertsch,Brussard,octup}. In the situation the eigenstates obtained for
$H_Q^{(-,-,-)}$  for the $K=0$ band are of
the form 
$$
\l|(\la_p , 0)(\la_n , 0)(\la_p + \la_n , 0)K=0,L,S=0,J=L\ran\;,
$$
formulas for the $Q(J)$ and $B(E2)$'s generated by our choice of $T^{E2}$ are given by,
\be
\barr{l}
Q((\la 0)J) = -\dis\f{J}{2J+3} (2\la +3)\,X_{eff} b^2\;,\\ 
B(E2; (\la 0)J \rightarrow J-2) = \dis\frac{5}{16\pi} \l\{
\dis\frac{6J(J-1)(\la -J+2)(\la +J+1)}{(2J-1)(2J+1)}\r\}(X_{eff})^2 b^4\;; \\
X=\dis\frac{e^p_{eff}\l(\la_p^2 + 3\la_p +\la_p \la_n
\r) + e^n_{eff}\l(\la_n^2 + 3\la_n +\la_n \la_p \r)}{
\l(\la^2 +3\la\r)}\;,\;\; \la=\la_p + \la_n \;.
\earr \label{eq.epj-9}
\ee
These formulas are not valid for $H_Q^{(\baa)} \neq H_Q^{(-,-,-)}$. Thus, numerical SM and DSM results are needed for the analysis of the eight algebras. 

\subsection{Results for $(sdgi)^{6p}$ system}

In the example with 6 protons in $\eta=6$ shell, SM matrix dimension in the
$m$-scheme $\sim 18 \times 10^5$. For this system, the leading $SU(3)$ irrep is
$(30,0)$ with $S=0$ and $T=3$ giving clearly $J=L$. It is seen that the SM calculations
reproduce the $SU(3)$ results  $E_{gs}=-990$ and  excitation energies $0.75
J(J+1)$ for all the eight $H$'s (note that the $H_Q$ matrix elements
are unit less and hence $E$ are unit less - in practical applications we have to put back appropriately the unit MeV). Thus, all the eight $H_Q$'s give $SU(3)$
symmetry. Though the energy spectra are same, the wave functions of the yrast $J$
states are different. This is established by calculating $Q(J)$ and $B(E2)$'s
for the ground band members. Choosing $e_{eff}^p=1e$ and $b^2=4.644 fm^2$, the
calculated results for $Q(2^+_1)$ in $e\,fm^2$ unit are $-84$, $-51$, $-19$,
$14$, $30$, $-3$, $-35$ and $-68$ for $(\baa) = (-,-,-)$, $(-,-,+)$, $(-,+,-)$,
$(-,+,+)$, $(+,+,+)$, $(+,+,-)$, $(+,-,+)$, $(+,-,-)$ respectively. Similarly,
for $Q(4^+_1)$ they are $-106$, $-65$, $-25$, $16$, $38$, $-3$, $-43$,  $-84$
respectively. The $B(E2)$'s also follow the same trend and for example $B(E2;
2^+_1 \rightarrow 0^+_1)$ values in $e^2 fm^4$ unit are 1700, 624, 82, 52, 217,
2, 306 and 1138 for the eight $H_Q$'s respectively. Thus, out of the eight
$SU(3)$ algebras, two of them generate oblate shape and the remaining six
prolate shape. Out of these six, one of them generates very small quadrupole
moments and $B(E2)$ values. All these SM results for the energies of the ground $K=0$ band, $Q(J)$'s and $B(E2)$'s are also well reproduced (within 5\% difference) by DSM
using a single intrinsic state as in the $(sdg)$ examples presented in
\cite{Ko-19}. The HF sp spectrum  for the $(sdg)^{6p}$ system is essentially same as the one shown in Fig. 1 except for a scale factor for the sp energies and the single intrinsic state employed in DSM calculations corresponds to 2 protons each in the lowest two $k=1/2$ sp levels and the lowest $k=3/2$ sp level shown in Fig. 1. Let us add that the lowest intrinsic state gives the
intrinsic quadrupole moments (in units of $b^2$) to be $60$, $35$, $15$, $-9$, 
$-21$, $3$, $23$ and $48$ for $(\baa) = (-,-,-)$, $(-,-,+)$, $(-,+,-)$, $(-,+,+)$,
$(+,+,+)$, $(+,+,-)$, $(+,-,+)$, $(+,-,-)$ respectively; the quadrupole operator is given by Eq. (\ref{eq.e2op}) with $e^p_{eff}=1$ and $e^n_{eff}=1$.
Now, we will consider the larger space example of $(sdgi)^{6p,6n}$
where SM calculations are not feasible and DSM gives the results.

\begin{table}

\caption{Deformed shell model results for quadrupole moments  $Q(J)$ (in $e\,fm^2$ unit) and $B(E2;
J \rightarrow J-2)$ values (in $e^2\,fm^4$ unit) for the ground $K=0^+$ band members for a system of 6
protons and 6 neutrons (with $T=0$) in  $\eta=6$ shell. Results are given for   the eight $(-1/4)Q^2(\baa) \cdot Q^2(\baa)$
Hamiltonians. In the table, $Q$ denotes $Q(J)$ and $B(E2)$ denotes $B(E2; J \rightarrow J-2)$.}
\label{table2}

\begin{tabular}{ccccccc}
\hline\noalign{\smallskip}
$\baa$ & $\;\;\;\;\;$ & \multicolumn{5}{c}{$J$} \\
& & 2 & 4 & 6 & 8 & 10 \\
\noalign{\smallskip}\hline\noalign{\smallskip}
$(-,-,-)$ & $Q$ & $-159$ & $-202$ & $-222$ &$-234$ & $-241$ \\
& $B(E2)$ & $6482$ & $9236$ & $10123$ & $10523$ & $10711$ \\ 
$(-,-,+)$ & $Q$ & $-95$ & $-121$ & $-133$ & $-141$ & $-146$ \\
& $B(E2)$ & $2309$ & $3263$ & $3525$ & $3587$ & $3550$ \\ 
$(-,+,-)$ & $Q$ & $-39$ & $-51$ & $-57$ & $-62$ & $-67$ \\
& $B(E2)$ & $391$ & $575$ & $667$ & $750$ & $840$ \\ 
$(-,+,+)$ & $Q$ & $25$ & $31$ & $32$ & $31$ & $29$ \\
& $B(E2)$ & $161$ & $225$ & $238$ & $235$ & $223$ \\ 
$(+,+,+)$ & $Q$ & $56$ & $72$ & $78$ & $82$ & $84$ \\
& $B(E2)$ & $822$ & $1172$ & $1286$ & $1340$ & $1368$ \\ 
$(+,+,-)$ & $Q$ & $-7$ & $-9$ & $-10$  &$-11$ & $-12$ \\
& $B(E2)$ & $14$ & $22$ & $29$ & $37$ & $48$ \\ 
$(+,-,+)$ & $Q$ & $-63$ & $-80$ & $-87$ & $-90$ & $-91$ \\
& $B(E2)$ & $1029$ & $1435$ & $1515$ & $1490$ & $1409$ \\ 
$(+,-,-)$ & $Q$ & $-127$ & $-161$ & $-175$ &$-183$ & $-186$ \\
& $B(E2)$ & $4164$ & $5907$ & $6426$ & $6609$ & $6633$ \\ 
\noalign{\smallskip}\hline
\end{tabular}
\end{table}   

\subsection{Results for $(sdgi)^{6p,6n}$ system} 

Carrying out DSM calculations for the $(sdgi)^{(6p,6n)T=0}$ system using the
eight $H_Q$'s, it is found that all of them generate the same  HF sp spectrum as
shown in Fig. 1. The lowest intrinsic state shown in the figure gives the
intrinsic quadrupole moments (in units of $b^2$) to be $120$, $70$, $34$, $-16$, 
$-43$, $8$, $43$, $94$ for $(\baa) = (-,-,-)$, $(-,-,+)$, $(-,+,-)$, $(-,+,+)$,
$(+,+,+)$, $(+,+,-)$, $(+,-,+)$, $(+,-,-)$ respectively. These are obtained
using  the quadrupole operator defined by Eq. (\ref{eq.e2op}) with $e^p_{eff}=1$ and $e^n_{eff}=1$. The intrinsic quadrupole
moments show that, again as in the $sdgi$IBM, in SM  also two of the eight
$SU(3)$ algebras generate oblate shape and rest of the six generate prolate
shape. Out of these six, one of them generates very small quadrupole moment. 
Going further, after
angular momentum projection from the lowest intrinsic state shown in the figure,
the ground state energy and the excited state energies of the yrast levels are
found to be within 1\% of the exact $SU(3)$ results (DSM generates yrast states with $S=0$ and $J=L$). Note that the $SU(3)$ irrep generating the ground $K=0$ band is $(60,0)$ giving 
$E(J=L)=-3780 + 0.75 J(J+1)$ for the yrast $0^+$, $2^+$, $4^+$, $\ldots$ levels. Turning to  $Q(J)$ and
$B(E2)$'s, in the calculations used are $e^p_{eff}=1.5e$, $e_{eff}^n=0.5e$ and
$b^2=4.644\,fm^2$. Here, Eq. (10) applies for the states from $H_Q^{(-,-,-)}$; note that $(\la_p,\mu_p)=(30,0)$ and $(\la_n,\mu_n)=(30,0)$. For $\baa=(-,-,-)$, the DSM results shown in Table 2 agree with the $SU(3)$ formulas to
within 3\%; However, the results from the other seven $H_Q$'s are quite
different as in the previous $(sdgi)^{6p}$ example.  Again, it is seen from
Table 2 that the results for $Q(J)$ and $B(E2)$'s from $H_Q^{(-,-,-)}$ and
$H_Q^{(+,-,-)}$ are strong and the $B(E2)$'s from $H_Q^{(+,+,-)}$ are much
smaller in magnitude. Moreover, six of them generate prolate shape and two of
them oblate shape as in the previous examples. Thus, the results in Tables 1 and
2 give the generic result that out of the eight $SU(3)$ algebras in the $(sdgi)$
space, six will give prolate and two oblate shape and in addition, one of them [$(+,+,-)$]
gives very small quadrupole moments. For further elucidating the difference between the eight $SU(3)$ algebras, we show in Table 3 the sp wave functions for the lowest sp state in Fig. 1 (this is same for protons and neutrons). With  sp $j$ values taking $1/2$ to $13/2$, the structure of the sp wave function is
\be
\l| k_r(\baa)\ran = \dis\sum_j (-1)^{\phi_j(\baa)} C_{k_r}^j \l| j k_r \ran\;,\;\;\; \mbox{with}\;\; C_{k_r}^j \ge 0\;.
\label{eq.epj-10}
\ee
Most important point here is that for a given $k_r$, the $C_{k_r}^j$ do not depend on $\baa$. Thus, the sp wave functions differ only in $(-1)^{\phi_j}$. Now, we will consider the $(sdgi)^{12p,6n}$
example.

\begin{table}

\caption{Phases $(-1)^{\phi_j(\baa)}$ in Eq. (\ref{eq.epj-10}) for the lowest Hartree-Fock sp state $\l|1/2_1\ran$ (see Fig. 1)  for the eight $(-1/4)Q^2(\baa) \cdot Q^2(\baa)$ Hamiltonians in $(sdgi)$ space. The 
$C_{1/2_1}^j$ for the seven sp states (in the order given in the table) are $0.366$, $0.430$, $0.527$, $0.380$, $0.425$, $0.191$ and $0.207$ respectively.}
\label{table3}
\begin{tabular}{ccccccccc}
\hline\noalign{\smallskip}
$\l|k_r\ran$ & $\baa$ & \multicolumn{7}{c}{$j$} \\
& & $\f{1}{2}$ & $\f{3}{2}$ & $\f{5}{2}$ & $\f{7}{2}$ & $\f{9}{2}$ & $\f{11}{2}$ & $\f{13}{2}$ \\
\noalign{\smallskip}\hline\noalign{\smallskip}
$1/2_1$ & $(-,-,-)$ & $-$ & $-$ & $+$ & $+$ & $-$ & $-$ & $+$ \\
& $(-,-,+)$ & $-$ & $-$ & $+$ & $+$ & $-$ & $+$ & $-$ \\ 
& $(-,+,-)$ & $-$ & $-$ & $+$ & $-$ & $+$ & $+$ & $-$ \\
& $(-,+,+)$ & $-$ & $-$ & $+$ & $-$ & $+$ & $-$ & $+$ \\
& $(+,-,-)$ & $+$ & $-$ & $+$ & $+$ & $-$ & $-$ & $+$ \\
& $(+,-,+)$ & $+$ & $-$ & $+$ & $+$ & $-$ & $+$ & $-$ \\
& $(+,+,-)$ & $+$ & $-$ & $+$ & $-$ & $+$ & $+$ & $-$ \\
& $(+,+,+)$ & $+$ & $-$ & $+$ & $-$ & $+$ & $-$ & $+$ \\
\noalign{\smallskip}\hline
\end{tabular}
\end{table}
\begin{figure}\sidecaption
\resizebox{0.5\columnwidth}{!}{
\includegraphics{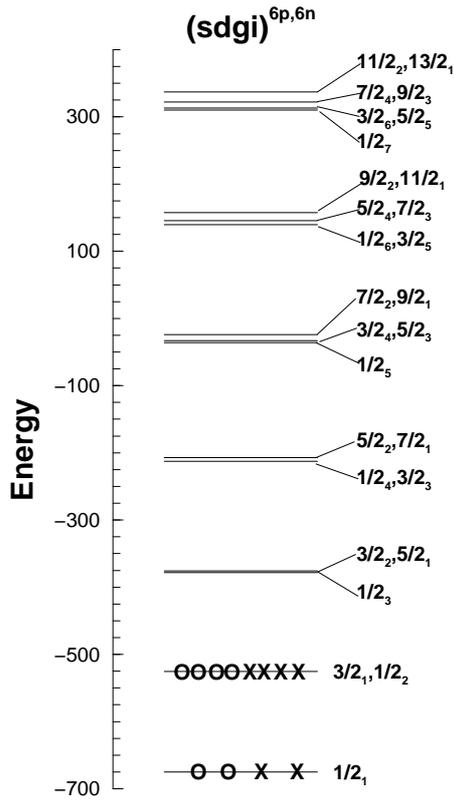} }
\caption{Hartree-Fock sp spectrum (it is same for both protons and neutrons) and
the lowest intrinsic state for the  $(sdgi)^{6p,6n}$ system generated by the
eight $H_Q$ operators.  In the figure, the symbol O denotes protons and
$\times$ denotes neutrons.  The spectrum is same for all  the eight
Hamiltonians although the sp wave functions are different. See text for further
details.}
\label{fig-1}  
\end{figure}
\begin{figure}\sidecaption
\resizebox{0.5\columnwidth}{!}{
\includegraphics{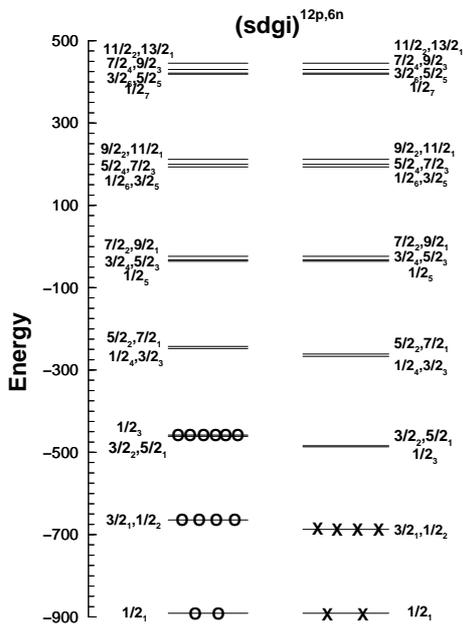} }
\caption{Hartree-Fock sp spectrum for
the lowest intrinsic state for the  $(sdgi)^{12p,6n}$ system generated by the
eight $H_Q$ operators. In the figure, the symbol O denotes protons and
$\times$ denotes neutrons.  The spectrum is same for all  the eight
Hamiltonians although the sp wave functions are different. See text for further
details.}
\label{fig-2}
\end{figure}

\subsection{Results for $(sdgi)^{12p,6n}$ system}

In our third example, DSM calculations for the $(sdgi)^{(12p,6n)T=3}$ system are carried out using the
eight $H_Q$'s. As in the previous examples, it is found that all of them generate the same  HF sp spectrum as
shown in Fig. 2. The lowest intrinsic state shown in the figure generates the ground $K=0$ band and its $SU(3)$ structure (same for all eight $H_Q$'s) is
$$
\l|(sdgi)^{12p} (48,0),\; (sdgi)^{6n} (30,0): (78,0)K=0,L;S=0;J=L;T=3\ran\;.
$$
For the lowest intrinsic state, the
intrinsic quadrupole moments (in units of $b^2$) are $156$, $70$, $71$, $-16$, 
$-51$, $35$, $34$, $121$ for $(\baa) = (-,-,-)$, $(-,-,+)$, $(-,+,-)$, $(-,+,+)$,
$(+,+,+)$, $(+,+,-)$, $(+,-,+)$, $(+,-,-)$ respectively. These are obtained again by
using  Eq. (\ref{eq.e2op}) with $e^p_{eff}=1$ and $e^n_{eff}=1$. The intrinsic quadrupole
moments show that, in this SM example also two of the eight
$SU(3)$ algebras generate oblate shape and rest of the six generate prolate
shape. Out of the six prolate examples, one of them generates very small quadrupole moments. 
Going further, after
angular momentum projection from the lowest intrinsic state shown in the figure,
the ground state energy and the excited state energies of the yrast levels are
found to be within 1\% of the exact $SU(3)$ result
$E(J=L)=-6318 + 0.75 J(J+1)$ for the yrast $0^+$, $2^+$, $4^+$, $\ldots$ levels. Turning to  $Q(J)$ and
$B(E2)$'s, in the calculations used is $T^{E2}$ in Eq. (\ref{eq.e2op}) with $e^p_{eff}=1.5e$, $e_{eff}^n=0.5e$ and
$b^2=4.61\,fm^2$. The DSM results are shown in Table 4. It is easy to see, by applying Eq. (10), that the DSM results for
$\baa=(-,-,-)$ agree with the exact $SU(3)$ formulas.
However, the results from the other seven $H_Q$'s are quite
different as in the previous $(sdgi)$ example.  Again, it is seen from
Table 4 that the results for $Q(J)$ and $B(E2)$'s from $H_Q^{(-,-,-)}$ and
$H_Q^{(+,-,-)}$ are strong and the $B(E2)$'s from $H_Q^{(-,+,+)}$ are much
smaller in magnitude. Moreover, six of them generate prolate shape and two of
them oblate shape as in the previous examples. Thus, the numerics in Tables 1, 2 and 4 give the generic result that out of the eight $SU(3)$ algebras in the $(sdgi)$
space, six will give prolate and two oblate shape and in addition, one of them 
gives small quadrupole moments. This need to be coupled with the fact that all of them generate the same spectrum. 

\begin{table}

\caption{Deformed shell model results for quadrupole moments  $Q(J)$ (in $e\,fm^2$ unit) and $B(E2;
J \rightarrow J-2)$ values (in $e^2\,fm^4$ unit) for the ground $K=0^+$ band members for a system of 12
protons and 6 neutrons (with $T=3$) in  $\eta=6$ shell. Results are given for the  the eight $(-1/4)Q^2(\baa) \cdot Q^2(\baa)$
Hamiltonians. In the table, $Q$ denotes $Q(J)$ and $B(E2)$ denotes $B(E2; J \rightarrow J-2)$.}
\label{table4}

\begin{tabular}{ccccccc}
\hline\noalign{\smallskip}
$\baa$ & $\;\;\;\;\;$ & \multicolumn{5}{c}{$J$} \\
& & 2 & 4 & 6 & 8 & 10 \\
\noalign{\smallskip}\hline\noalign{\smallskip}
$(-,-,-)$ & $Q$ & $-234$ & $-297$ & $-327$ &$-344$ & $-355$ \\
& $B(E2)$ & $13288$ & $18952$ & $20815$ & $21699$ & $22167$ \\ 
$(-,-,+)$ & $Q$ & $-92$ & $-118$ & $-130$ & $-138$ & $-144$ \\
& $B(E2)$ & $2065$ & $2953$ & $3256$ & $3414$ & $3514$ \\ 
$(-,+,-)$ & $Q$ & $-120$ & $-152$ & $-167$ & $-176$ & $-181$ \\
& $B(E2)$ & $3501$ & $4988$ & $5467$ & $5682$ & $5782$ \\ 
$(-,+,+)$ & $Q$ & $22$ & $27$ & $30$ & $31$ & $31$ \\
& $B(E2)$ & $114$ & $161$ & $176$ & $182$ & $184$ \\ 
$(+,+,+)$ & $Q$ & $76$ & $96$ & $104$ & $109$ & $110$ \\
& $B(E2)$ & $1395$ & $1988$ & $2180$ & $2267$ & $2309$ \\ 
$(+,+,-)$ & $Q$ & $-66$ & $-84$ & $-93$ & $-98$ & $-101$ \\
& $B(E2)$ & $1055$ & $1501$ & $1642$ & $1703$ & $1727$ \\ 
$(+,-,+)$ & $Q$ & $-38$ & $-49$ & $-55$ & $-60$ & $-64$ \\
& $B(E2)$ & $352$ & $504$ & $559$ & $591$ & $615$ \\ 
$(+,-,-)$ & $Q$ & $-180$ & $-229$ & $-252$& $-266$ & $-276$ \\
& $B(E2)$ & $7846$ & $11191$ & $12290$ & $12812$ & $13087$ \\ 
\noalign{\smallskip}\hline
\end{tabular}
\end{table}   

\section{Conclusions}

Multiple $SU(3)$ algebras appearing in both the shell model and the interacting
boson model opened a  new paradigm in the applications of $SU(3)$ symmetry in
nuclei.  The $sdgi$IBM and SM examples presented  in Sections 2-4 show that the
eight $SU(3)$ algebras in the $(sdgi)$ space of IBM and SM exhibit quite
different properties with regard to the quadrupole collectivity as brought out
by the quadrupole moments $Q(J)$ and $B(E2)$'s in the ground $K=0$ bands. Six of
them generate prolate shape, two oblate shape and in the six prolate, one of
them generates small quadrupole moments.  However, they all generate the
same rotational spectra. Thus, with multiple $SU(3)$ algebras it is possible to
have rotational spectra with strong quadrupole collectivity and also  rotational
spectra with weak quadrupole collectivity. In addition, some of them give
prolate and other oblate shapes. All these conclusions are consistent with the
earlier $sdg$ space results \cite{Ko-19} and therefore establish that these are generic results valid both in shell model and interacting boson model spaces.  

Further understanding of multiple
$SU(3)$ algebras in $sdgi$IBM spaces will follow by analyzing the structure of
the low-lying  $\gamma$ and  $\beta$ bands. It is also of interest to analyze
multiple $SU(3)$ algebras in IBM-2 in $(sdg)$ and $(sdgi)$ spaces. In some of these IBM studies, large $N$
results  given in \cite{KuMo-1,KuMo-2} will be useful. Note that in IBM-2, it is possible to consider multiple $SU(3)$ algebras in proton bosons space and neutron boson space separately, leading to a much larger class of $SU(3)$ algebras. Turning to SM
spaces,  SM and DSM analysis of systems with lowest $SU(3)$ irrep of the type
$(\la \mu)$ with $\mu \neq 0$  will be important. Here,  $SU(3) \supset SO(3)$
integrity basis operators that are 3 and 4-body are needed as
demonstrated for example in \cite{jpd-1}. Secondly, the $H$'s defined
by multiple $SU(3)$ algebras in $sdg$ and higher spaces are expected to be important in QPT studies (example with $sd$IBM is known \cite{RMP}) and this will be explored in a future publication. Also, as in IBM-2, in the shell model studies of heavy nuclei with protons and neutrons in different oscillator shells, there will be multiple $SU(3)$ algebras in the SM space for protons and in the SM space for neutrons separately. Combining these proton and neutron $SU(3)$ algebras will again lead to a much larger class of multiple $SU(3)$ algebras in SM. These will be investigated in a future publication. 
Let us add that, multiple $SU(3)$ algebras discussed here combined with the multiple
pairing algebras in SM and IBM spaces \cite{Ko-17,Ko-so8} will generate multiple pairing plus quadrupole-quadrupole ($P + Q \cdot Q$) Hamiltonians. These are expected to give new insights
into the structures generated by ($P + Q \cdot Q$) Hamiltonians in nuclei; see \cite{Kalin} for new interest in the studies using $P + Q \cdot Q$ Hamiltonians. Finally, it is important to examine experimental data testing the results of multiple $SU(3)$ algebras in $sdg$ and $sdgi$ spaces and this is postponed to a future study. 

\begin{acknowledgement}

Thanks are due to N.D. Chavda for some computational help.
RS is thankful to SERB of DST, Government of India for financial support and
PCS acknowledges a research grant from SERB (India), CRG/ 2019/000556.

\end{acknowledgement}

\appendix
\renewcommand{\theequation}{A\arabic{equation}}
\setcounter{equation}{0}   

\begin{center}
Appendix-A
\end{center}

Methods for obtaining the spe and TBME for the quadrupole-quadrupole interaction operator $Q^2(\baa) \cdot Q^2(\baa)$ (for all
phase choices $\baa$) are well known \cite{Brussard}. In
order to derive the formulas for the spe and TBME, we begin with Eq. (\ref{eq.epj-8})
and drop the factor '2'. Also, we do not show $\baa$ in $Q^2_q(\baa)$ when there is no confusion. For a many particle system, 
\be
Q \cdot Q = \sum_{i=1}^m Q(i) \cdot Q(i) + 2 \sum_{i<k=1}^m Q(i) \cdot 
Q(k)
\label{eq.aap-1}
\ee
where $i$ and $k$ are particle indices and $m$ is number of particles.  The
first sum generates spe and the second term TBME. Given the shell model sp $(n
\ell j)$-orbits (note that the oscillator shell number $\eta=2n+\ell$), matrix
elements of $Q(1) \cdot Q(2)$ in the two-particle  antisymmetric states (called
a.s.m.)  can be written in terms of the matrix elements in the two-particle
non-antisymmetric states (called n.a.s.m.) as,
\be
\barr{l}
\lan (j_a j_b)JT \mid Q(1) \cdot Q(2) \mid(j_c j_d)JT\ran_{a.s.m.} = \dis\f{1}
{\dis\sqrt{\l(1+\delta_{ab}\r) \l(1+\delta_{cd}\r)}} \\
\\
\times \; \l[\lan (j_a j_b)JT \mid Q(1) \cdot Q(2) \mid(j_c j_d)JT
\ran_{n.a.s.m.} \r. \\
\\
\l. +
(-1)^{J+T-j_c -j_d} \lan (j_a j_b)JT \mid Q(1) \cdot Q(2) \mid(j_d j_c)JT\ran_{
n.a.s.m.}\r]\;.
\earr\label{eq.aap-2}
\ee
Using angular momentum algebra it is easy to recognize that,
\be
\barr{l}
\lan (j_a j_b)JT \mid Q(1) \cdot Q(2) \mid(j_c j_d)JT\ran_{n.a.s.m.} = 
(-1)^{j_b + j_c +J} \l\{\barr{ccc} j_a & j_b & J\\ j_d & j_c & 2\earr\r\} \\
\times\;\lan j_a \mid\mid Q \mid\mid j_c\ran\,\lan j_b \mid\mid Q \mid\mid 
j_d\ran\;.
\earr \label{eq.aap-3}
\ee
The reduced matrix elements $\lan \mid\mid Q \mid\mid \ran$ are given by, 
\be
\barr{l}
\lan \eta, \ell_f, j_f \mid\mid Q^2(\baa) \mid\mid \eta, \ell_i, j_i\ran = 
(-1)^{\ell_f + \frac{1}{2} + j_i + 2} \\
\times\;\dis\sqrt{5 (2j_i +1)(2j_f +1)}\;\l\{\barr{ccc}\ell_f & j_f &
\frac{1}{2} \\ j_i & \ell_i & 2 \earr\r\}\;t_{\ell_f , \ell_i}(\baa) \;.
\earr \label{eq.aap-4}
\ee
Combining Eqs. (\ref{eq.aap-3}) and (\ref{eq.aap-4}) with Eq. (\ref{eq.aap-2}) and 
Eq. (\ref{eq.aap-1}) will give the TBME of the $Q^2(\baa) \cdot Q^2(\baa)$ 
operator. The spe $\epsilon^{\baa}_{\ell j}$ of the $Q^2(\baa) \cdot Q^2(\baa)$
are simply given by
\be
\epsilon^{\baa}_{\ell j} = \dis\frac{5}{2\ell +1}\; \dis\sum_{\ell^\prime} 
\l|t_{\ell \ell^\prime}(\baa)\r|^2\;.
\ee
An important property is, 
\be
-\dis\f{1}{4}\,Q^2(\baa) \cdot Q^2(\baa) = \cc_2(SU^{\baa}(3)) + 
\dis\f{3}{4} L \cdot L \;.
\label{eq.aap-5}
\ee

\end{document}